\documentclass{iaus}
\usepackage{graphicx}
\usepackage{amssymb}
\usepackage{amsmath}

\newcommand{\MBH}{M_{\text{BH}}}
\newcommand{\MDM}{M_{\text{DM}}}

\newcommand{\vc}{{v_{\text{c}}}}

\newcommand{\vcsigma}{$\vc$\,--\,$\sigma$}
\newcommand{\MBHMDM}{$\MBH$\,--\,$\MDM$}
\newcommand{\MBHsigma}{$\MBH$\,--\,$\sigma$}
\newcommand{\MBHvc}{$\MBH$\,--\,$\vc$}

\newcommand{\kms}{{\text{km\,s$^{-1}$}}}
\newcommand{\Msun}{{\text{M$_\odot$}}}
\newcommand{\vunit}{v_0}

\title%
[The Interplay among Black Holes, Stars and ISM in Galactic Nuclei]%
{\vspace*{5mm}% 
Observational evidence for a connection between SMBHs and dark matter haloes}

\author[M. Baes, H. Dejonghe, P. Buyle, L. Ferrarese \& G. Gentile]{%
Maarten Baes$^{1,2}$,
Herwig Dejonghe$^1$,
Pieter Buyle$^1$,
Laura Ferrarese$^3$,
Gianfranco Gentile$^{4,5}$}

\affiliation{%
$^1$Sterrenkundig Observatorium, Universiteit Gent, \\
Krijgslaan 281 S9, B-9000 Gent, Belgium, email: maarten.baes@ugent.be\\[\affilskip]
$^2$Department of Physics and Astronomy, Cardiff University, \\
5 The Parade, Cardiff CF24\,3YB, Wales, UK\\[\affilskip]
$^3$Rutgers University, New Brunswick, NJ 08854, USA\\[\affilskip]
$^4$Radioastronomisches Institut der Universit\"at Bonn, \\
Auf dem H\"ugel 71D, D-53121 Bonn, Germany\\[\affilskip]
$^5$SISSA, via Beirut 4, 34014 Trieste, Italy%
}

\pubyear{2004}
\volume{222}
\pagerange{1--8}
\date{?? and in revised form ??}
\jname{The Interplay among Black Holes, Stars and ISM \\in Galactic Nuclei}
\editors{Th. Storchi Bergmann, L.C. Ho \& H.R. Schmitt, eds.}
\begin{document}

\maketitle

\begin{abstract}
We investigate the relation between circular velocity $\vc$ and bulge
velocity dispersion $\sigma$ in spiral galaxies, based on literature
data and new spectroscopic observations. We find a strong, nearly
linear \vcsigma\ correlation with a negligible intrinsic scatter, and
a striking agreement with the corresponding relation for elliptical
galaxies. The least massive galaxies ($\sigma<80~\kms$) significantly
deviate from this relation. We combine this \vcsigma\ correlation with
the well-known \MBHsigma\ relation to obtain a tight correlation
between circular velocity and supermassive black hole mass, and
interpret this as observational evidence for a close link between
supermassive black holes and the dark matter haloes in which they
presumably formed. Apart from being an important ingredient for
theoretical models of galaxy formation and evolution, the relation
between $M_{\text{BH}}$ and circular velocity has the potential to
become an important practical tool in estimating supermassive black
hole masses in spiral galaxies.
\end{abstract}

\firstsection % if your document starts with a section,
              % remove some space above using this command.
\section{Introduction}

The existence of supermassive black holes (SMBHs) in the nuclei of
galaxies has been suspected for almost half a decade, as accretion
onto SMBHs seemed the only logical explanation for the existence of
quasars. HST observations have provided evidence that SMBHs with
masses ranging from $10^6$ to $10^9$~\Msun\ are present in the centre
of a few dozens of nearby (quiescent) galaxies. Be this sufficient
evidence for the existence of SMBHs, we can now tackle more
fundamental questions concerning their formation and evolution. An
obvious way to proceed is the study of the relation between SMBHs and
the galaxies that host them. It was found that black hole masses are
correlated with parameters of the hot stellar components of their host
galaxies. The tight \MBHsigma\ relation (Gebhardt et al.~2000;
Ferrarese \& Merritt~2000) is now the preferred paradigm to study SMBH
demographics in galactic nuclei.

This apparently tight link between bulges and SMBHs reflects an
important ingredient that should be reproduced (and thus hopefully
explained) by theoretical models of galaxy formation. In fact, the
tightness of the \MBHsigma\ correlation is somewhat surprising. In
most of the state-of-the-art models, the total galaxy mass (or dark
matter mass $\MDM$), rather than the bulge mass, plays a fundamental
role in shaping the SMBHs. A close correlation could therefore be
expected between $\MBH$ and $\MDM$, rather than between $\MBH$ and the
bulge properties. Establishing whether the \MBHsigma\ or the \MBHMDM\
relation reflects the fundamental mode by which SMBHs form and evolve
will ultimately rely on a comparison of the intrinsic scatter of the
two correlations.

Unfortunately, a direct observational characterization of the \MBHMDM\
relation is currently impossible. Ferrarese~(2002b) first argued that
a correlation between $\MBH$ and $\MDM$ should be reflected in an
\MBHvc\ correlation, where $\vc$ is the circular velocity in the flat
part of the rotation curve of spiral galaxies. Indeed, in most of the
state-of-the-art galaxy formation models, there is a one-to-one
correspondence between the circular velocity and the mass of the dark
matter halo. Unfortunately, there are (presently) only a handful of
spiral galaxies with secure SMBH masses, and only two of them have a
well-measured extended rotation curve. A way to avoid this problem is
to adopt the tight \MBHsigma\ correlation in order to estimate black
hole masses in a larger sample of galaxies. A tight correlation
between SMBH mass and dark matter halo mass should thus appear in the
form of a correlation between central velocity dispersion and circular
velocity. Ferrarese~(2002b) presented a first attempt at establishing
such a correlation. Baes et al.~(2003) significantly improved on these
results by almost doubling the sample size. The present contribution
is focused on the latter results.

\section{Sample selection}

A simple measure for the circular velocity of galaxies is half of the
integrated line width from spatially unresolved H{\sc i} 21\,cm
measurements, corrected for inclination. Various authors have
recovered a nearly linear correlation between integrated line width
and bulge velocity dispersion (e.g.\ Whitmore \& Kirshner~1981;
Whittle~1992; Franx~1993). This correlation has significant scatter
and galaxy type could act as a third parameter in this
correlation. This can be due to the fact that the integrated line
width is not an accurate measure for the flat part of the rotation
curve, and hence of the total dark matter content. Therefore, we chose
to consider only those galaxies with an extended rotation curve
measured well beyond the optical radius, in order to reliably trace
the circular velocity in the flat part of the rotation curve. In order
to see the benefits of this approach, it is interesting to consider
the Tully-Fisher (TF) study of the Ursa Major cluster spiral galaxies
by Verheijen~(2001): the scatter in the TF relation strongly decreases
when he considers the flat part in the rotation curve instead of the
integrated line width.

We constructed a data set of 28 spiral galaxies with central velocity
dispersion data and a rotation curve measured beyond the optical
radius. For 16 spirals, the data could be retrieved from the
literature (see references in Ferrarese~2002b). For the remaining 12
galaxies, rotation curve data were available in the literature
(Palunas \& Williams 2000), and the velocity dispersions were measured
with the EFOSC2 instrument on the ESO 3.6m telescope. The total sample
of 28 spiral galaxies can be found in Baes et al.~(2003).

\begin{figure*}
\centering
\includegraphics[bb=48 307 550 534,width=0.9\textwidth,clip]{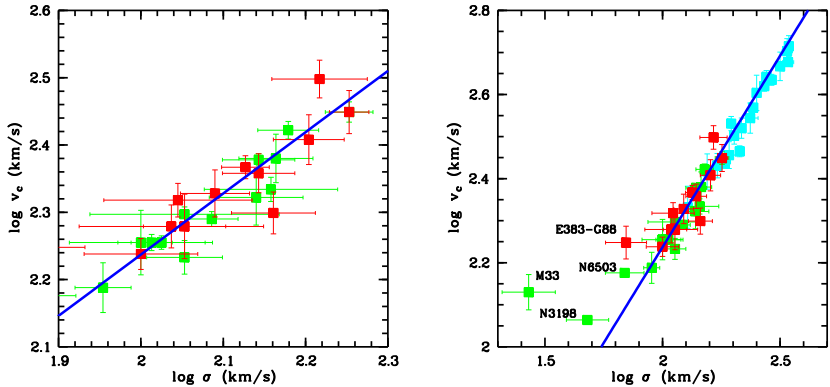}
\caption{
The correlation between the circular velocity $\vc$ and the central
velocity dispersion $\sigma$. The left plot shows the \vcsigma\
correlation for the 24 spiral galaxies with rotation curve beyond the
optical radius and a velocity dispersion $\sigma>80$~\kms. The data
points are from Ferrarese~(2002b) and Baes et al.~(2003)
respectively. The right plot zooms out and adds the four spiral
galaxies with $\sigma<80$~\kms and the elliptical galaxies from
Kronawitter et al.~(2000).}
\label{vcsigma.eps}
\end{figure*}

\section{The \vcsigma\ correlation}

In the left panel of figure~{\ref{vcsigma.eps}} we plot the circular
velocity versus the velocity dispersion for the 28 spiral galaxies in
our sample. For the 24 galaxies with a velocity dispersion greater
than about 80~\kms, there is a very tight correlation between $\vc$
and $\sigma$. We fitted a straight line to these data, taking into
account the errors on both quantities and obtained
\begin{displaymath}
	\log\left(\frac{\vc}{\vunit}\right) 
	= 
	(0.96 \pm 0.11) \log\left(\frac{\sigma}{\vunit}\right) 
	+ 
	(0.21 \pm 0.023),
\label{vcsigma2}
\end{displaymath}
where $\vunit=200\,\kms$. Two issues concerning this \vcsigma\
correlation deserve some special attention.

Firstly, the tightness of the correlation is astonishing: we find
$\chi_{\text{red}}^2 = 0.281$, corresponding to a goodness-of-fit of
99.9 per cent. The \vcsigma\ relation can hence be regarded as having
a negligible intrinsic scatter. Moreover, this correlation appears to
be robust: there are no significant outliers in the range
$\sigma>80~\kms$. The correlation appears to break down for galaxies
with dispersions below about 80~\kms\ however: all four galaxies with
$\sigma<80~\kms$ lie significantly above the correlation defined by
the more massive spirals. Interestingly, this is also the mass range
in which nearly all bulgeless spiral galaxies are located. New
observations are indispensable to understand the behaviour of the
\vcsigma\ correlation in the low mass regime. Moreover, all galaxies
in our sample are high surface brightness galaxies, and it presently
unclear how the \vcsigma\ relation behaves in diverse environments
(see also Pizzella et al.~2004).

Secondly, it is interesting to compare this correlation to a similar
one recently found for elliptical galaxies. Based on stellar dynamical
models for 20 round ellipticals constructed by Kronawitter et
al.~(2000), Gerhard et al.~(2001) discovered a very tight relation
between the central dispersion and the circular velocity (the circular
velocity curves of ellipticals were found to be flat to within 10 per
cent). Both the slope and zero-point of this correlation agree
amazingly well with the \vcsigma\ correlation of our spiral galaxy
sample (see right panel of figure~1). Gerhard et al.~(2001) argue that
a proportionality between $\sigma$ and $\vc$ can be expected for
ellipticals on the basis of their dynamical homology. For spiral
galaxies this proportionality cannot be explained by simple dynamical
arguments, as convincingly argued by Ferrarese~(2002b). Moreover, the
fact that both spiral and elliptical galaxies seem to obey exactly the
same correlation is absolutely striking.

\section{The correlation between $\MBH$ and $\vc$}

As both the \vcsigma\ and \MBHsigma\ correlations seem to hold over
the entire Hubble range, we can combine them to derive a correlation
between the circular velocity and SMBH mass. Although the slope of the
\MBHsigma\ relation is not well established (Tremaine et al.~2002;
Ferrarese~2002a), this little affects the conclusion that the
\vcsigma\ relation entails a connection between SMBH mass and the
large scale velocity of the host galaxy (and henceforth the mass of
the surrounding dark matter halo). For instance, using the
characterization of the \MBHsigma\ relation from Tremaine et
al.~(2002), we obtain
\begin{displaymath}
	\log\left(\frac{\MBH}{\Msun}\right)
	=
	(4.21 \pm 0.60) \log\left(\frac{\vc}{\vunit}\right)
	+
	(7.24 \pm 0.17).
\label{MBHvc}
\end{displaymath}
The correlation between SMBH mass and circular velocity is useful for
two different goals. Firstly, combined with other tight relations such
as the \MBHsigma\ relation and the TF relation, it clearly points at
an intimate interplay between the various galactic components (dark
matter, discs, bulges and SMBHs) and forms a strong test for galaxy
formation and evolution models. In particular, the \vcsigma\ relation
can be used to discriminate between various theoretical models of
galaxy formation. For example, if SMBHs form mainly through
coalescence of smaller black holes during galaxy mergers, a relation
$\MBH\propto v_{\text{c}}^3$ is expected, whereas theories in which
accretion and feedback are the main ingredients for black hole growth
prefer a $\MBH\propto v_{\text{c}}^5$ relation (e.g.\ Silk \&
Rees~1998; Haiman \& Loeb~1998; Haehnelt, Natarajan \& Rees~1998;
Kauffman \& Haehnelt~2000; Wyithe \& Loeb~2003; Di Matteo et
al.~2003).

Apart from being an ingredient in theoretical galaxy formation models,
the derived \MBHvc\ relation can also serve as a practical tool to
estimate the black hole masses in spiral galaxies. The most preferred
means of estimating $\MBH$ in galaxies is the \MBHsigma\
relation. Unfortunately, the number of spiral galaxies with reliable
velocity dispersion measurements is relatively small. Since extended
rotation curves have been measured for large samples of spiral
galaxies (mainly for use in TF studies), the \MBHvc\ relation has the
potential to become an important practical tool in estimating
supermassive black hole masses in spiral galaxies.

\end{document}